# Solving the non-relativistic electronic Schrödinger equation with manipulating the coupling strength parameter over the electron-electron Coulomb integrals

Sandor Kristyan

*Research Centre for Natural Sciences, Hungarian Academy of Sciences, Institute of Materials and Environmental Chemistry, Magyar tudósok körútja 2, Budapest H-1117, Hungary,*

Corresponding author: kristyan.sandor@ttk.mta.hu

**Abstract**. The non-relativistic electronic Hamiltonian, $H(a) = H_\nabla + H_{ne} + aH_{ee}$, extended with coupling strength parameter (a), allows to switch the electron-electron repulsion energy off and on. First, the easier a=0 case is solved and the solution of real (physical) a=1 case is generated thereafter from it to calculate the total electronic energy ($E_{total\ electr,K}$) mainly for ground state (K=0). This strategy is worked out with utilizing generalized Moller-Plesset (MP), square of Hamiltonian ($H^2$) and Configuration interactions (CI) devices. Applying standard eigensolver for Hamiltonian matrices (one or two times) buys off the needs of self-consistent field (SCF) convergence in this algorithm, along with providing the correction for basis set error and correlation effect. (SCF convergence is typically performed in the standard HF-SCF/basis/a=1 routine in today practice.)

## INTRODUCTION

The non-relativistic, spinless, fixed nuclear coordinate electronic Schrödinger equation (SE) for molecular systems containing M atoms and N electrons with nuclear configuration $\{\mathbf{R}_A, Z_A\}_{A=1}^M$ in free space is

$$(H_\nabla + H_{ne} + H_{ee})\Psi_K = E_{electr,K}\Psi_K \quad \text{and} \quad (H_\nabla + H_{ne})Y_k = e_{electr,k}Y_k. \tag{1}$$

Energy operators are the kinetic ($H_\nabla \equiv -\Sigma_{i=1}^N \nabla_i^2/2$), nuclear–electron attraction ($H_{ne} \equiv -\Sigma_{i=1}^N \Sigma_{A=1}^M Z_A R_{Ai}^{-1}$) and electron–electron repulsion ($H_{ee} \equiv \Sigma_{i=1}^N \Sigma_{j=i+1}^N r_{ij}^{-1}$). The left one in Eq.1 is the physical equation (a=1), and in the right the operator $H_{ee}$ is switched off (a=0, called TNRS since all electron-electron repulsions are totally switched off) serving as a mathematic precursor equation for the left one. The eigenfunctions ($\Psi_K$ and $Y_k$) are anti-symmetric (with respect to all spin-orbit electronic coordinates $\mathbf{x}_i \equiv (\mathbf{r}_i, s_i)$), well behaving and normalized. The one-electron density is $\rho_k(\mathbf{r}_1, a) = N\int y_K^* y_K ds_1 d\mathbf{x}_2 \ldots d\mathbf{x}_N$ for both $y_K = \Psi_K$ or $Y_k$. The $y_{K=k}(a=0) = Y_k$ has a single determinant form, while $y_K(a=1) = \Psi_K$, generally $y_K(a \neq 0)$, does not. $E_{electr,0}$(method) approximates $E_{electr,0}$ by a certain method, and $E_{total\ electr,K} = E_{electr,K} + V_{nn}$, where $V_{nn} \equiv \Sigma_{A=1}^M \Sigma_{B=A+1}^M Z_A Z_B / R_{AB}$ is the nucleus-nucleus repulsion term. K=k=0 are called ground states, and K, k >0 are the excited states.

The associative $H(a) = (H_\nabla + aH_{ee}) + H_{ne} = (H_\nabla + H_{ne}) + aH_{ee}$ property is elementary, however it suggests two ways for solution: In density functional theory (DFT) the nuclear frame independent ($H_\nabla + H_{ee}$) operators were separated from the nuclear frame determining $H_{ne}$, yielding the 1st Hohenberg-Kohn (HK) theorem [1-2] from $\{\Psi_0 \Leftrightarrow H_{ne}\}$ as $\rho_0 \Rightarrow \{N, Z_A, R_A\} \Rightarrow H \Rightarrow \Psi_0 \Rightarrow \{E_{electr,0}, \text{all properties}\}$. It provides that $\{\Psi_0 \Leftrightarrow H \Leftrightarrow H_\nabla + H_{ne} \Leftrightarrow Y_0\}$, finally, $\{Y_0 \Leftrightarrow H_{ne} \Leftrightarrow \Psi_0\}$ or $\{\rho_0(\mathbf{r}_1, a=0) \text{ or } Y_0 \Leftrightarrow \rho_0(\mathbf{r}_1, a=1) \text{ or } \Psi_0\}$. More importantly: $Y_0, \rho_0(\mathbf{r}_1, a=0)$ from $H_\nabla + H_{ne} \Leftrightarrow E_{electr,0}$ from $H_\nabla + H_{ne} + H_{ee}$. By the latter, the TNRS algorithm is based on the separation of $(H_\nabla + H_{ne})$ from $H_{ee}$ in Eq.1.

$S_0$ is a normalized single Slater determinant approx. for $\Psi_0$ left in Eq.1 via Hartree-Fock Self Consistent Field HF-SCF/basis/a=1 energy minimizing algorithm [1-3] for the functional $<S_0|H|S_0> \geq E_{electr,0} = <\Psi_0|H|\Psi_0>$ with constrained (orthonormalized) molecular orbitals (MO). The error $<\Psi_0|H|\Psi_0> - <S_0|H|S_0>$, negative by variation principle (VP) and $\approx 1\%$ magnitude with at least minimal basis set at near stationary points, known as correlation energy ($E_{corr}(a \neq 0, K=0) < 0$) plus basis set error (<0 by VP), as well as for k>0 the estimation is very weak. The HF-SCF/basis/a=0 algorithm solves the right one in Eq.1 for $Y_k$ for ground- (k=0) and excited states (k>0), with single Slater determinant (correct form) for $Y_k$ with basis set error (<0 by VP) only, (and $E_{corr}(a=0, k \geq 0) = 0$). Interestingly, the $S_0$ is very close to $Y_0$ in its linear combination of atomic orbital (LCAO) coefficients (on the same basis set level).

Moreover, calculating $Y_0$ in this way is not restricted to the vicinity of the stationary point, while if $a \neq 0$ (e.g. for $S_0(a=1)$) it does. The HF-SCF/basis/a=0 calculation is faster, more stable and less memory taxing in comparison to HF-SCF/basis/a=1. In Eq.1 the right one is $(H_\nabla + H_{ne})Y_k = \Sigma_{i=1}^N h_i Y_k = e_{electr,k} Y_k$, where $h_i \equiv -\nabla_i^2/2 - \Sigma_{A=1}^M Z_A R_{Ai}^{-1}$ is the one-electron operator, and both (using general "a") decompose to one-electron equations with linear dependence on "a" as $(h_i + aV_{ee,eff}(\mathbf{r}_i))\phi_i(\mathbf{r}_i) = \varepsilon_i \phi_i(\mathbf{r}_i)$, where $V_{ee,eff}$ is the effective potential from electron-electron repulsion and $\phi_i(\mathbf{r}_i)$ is the i[th] MO. (Sum of this one particle Fock operators yields the approx. [3] HF-SCF/basis/a=1 (generally a≠0) for the left in Eq.1 yielding $S_0$, called virtually non-interacting reference system (VNRS), $S_0$ is exact single determinant solution of this equation system (up to basis set error), but not of left in Eq.1; if $a \to 0$, it becomes the TNRS, $S_0 \to Y_0$ which is not only the exact single determinant solution of this equation system but also of the right in Eq.1.) Technically $\phi_i$ counts the MOs with the index i, so the notation is reducible from $(h_i, \phi_i(\mathbf{r}_i), \varepsilon_i)$ to $(h_1, \phi_i(\mathbf{r}_1), \varepsilon_i)$; a=0 makes it uncoupled and a≠0 makes it coupled (via $r_{ij}^{-1}$ terms). In fact

$$h_1 \phi_i \equiv (-\nabla_1^2/2 - \Sigma_{A=1}^M Z_A R_{A1}^{-1})\phi_i(\mathbf{r}_1) = \varepsilon_i \phi_i(\mathbf{r}_1) \qquad (2)$$

is a single eigenvalue equation providing orthonormality $\langle\phi_i|\phi_j\rangle = \delta_{ij}$ by its mathematical nature, (in HF-SCF/basis/a≠0 the orthonormality is enforced). Among eigenvalues $(\varepsilon_i, \phi_i(\mathbf{r}_1))$ for i= 0,1,2,…∞, the i=0 is the lowest lying state and it is the lowest lying MO for a=0 in Eq.1 in its k=0 ground state. The single Slater determinant $Y_0$ for a=0 in Eq.1 is accomplished for N electrons from the set $\{\phi_i(\mathbf{r}_1)\}$, in the same manner as in the basic HF-SCF theory for $S_0$. The electronic energy of the system is $e_{electr,k} = \Sigma_i n_i \varepsilon_i$, while $E_{electr,0}(HF-SCF/basis/a=1) \neq \Sigma_i n_i \varepsilon_i$, where $n_i$ is the population of the i[th] energy level: 0, 1 or 2, the lattermost is with opposite spins, and $\Sigma_{i=1} n_i = N$. Important: 1, set $\{\phi_i, \varepsilon_i\}$ with i=0,1,2,…(N-2)/2 or (N-1)/2 calculated by HF-SCF/basis/a is independent of N if a=0, but depends on N if a≠0, the HF-SCF/basis/a=0 procedure can be used to calculate higher $(\phi_i, \varepsilon_i)$ states by simply increasing the N ("virtual" N) with adequate basis sets for excited states, still having the lower states algebraically intact; 2, HF-SCF/basis/a=0 needs only two steps, more exactly one, after setting up an initial guess for LCAO parameters, the eigensolver yields the $Y_0$ in the next step, irrespectively of molecular size (mainly N), in fact HF-SCF convergence is not needed at all, only one step engensolving, while HF-SCF/basis/a=1 needs more and more convergence steps as N increases. Starting with initial LCAO parameters for HF-SCF/basis/a=1 and finishing the convergence, or starting with LCAO from a converged one step HF-SCF/basis/a=0 and finishing the convergence for a=1, the final $E_{electr,0}$(HF-SCF/basis/a=1) and LCAO parameters will be strictly the same via the VP. For simplicity below, H without argument means H(a=1).

## CALCULATING GROUND AND EXCITED STATES WITH a=0

**Spin states in TNRS (a=0)** in Eq.1: H(a) does not contain spin coordinates ⇒ for $S_{op}^2$ and $S_{op,z}$ total spin operators

$$[H_\nabla + H_{ne} + aH_{ee}, S_{op}^2 \text{ or } S_{op,z}] = 0, \qquad (3)$$

i.e. commute with it ⇒ $y_K(a)$ are also eigenfunctions [3] of these. Particularly, if a=0 then

$$S_{op}^2 Y_k = S(S+1)Y_k \quad \text{and} \quad S_{op,z} Y_k = M_S Y_k \quad \text{with} \quad S \equiv \Sigma_{i=1}^N s_i, \qquad (4)$$

where S and $M_S$ are the total spin quantum number and its z component, the 2S+1 is the multiplicity. For the first, the single determinants are simple cases, particularly $Y_0$, because for a closed shell $Y_k$, the $S_{op}^2 Y_k = 0$ is a pure singlet [3], see Eq.5, but an open shell $Y_k$ is generally not eigenfunction of $S_{op}^2$ [3], except when all the open shell electrons have parallel spins, see Eq.6. Generally, if excitation from single determinant $S_0$ or $Y_k$ (using e.g. LUMO) for CI is not a "pure" spin state single determinant (e.g. $|(\alpha\phi_0)(\mathbf{r}_1), (\alpha\phi_i)(\mathbf{r}_2), (\beta\phi_i)(\mathbf{r}_3)\rangle$ is "pure" if i=0 or 1, not if 0≠i≠1), spin adapted configuration can be formed by taking appropriate linear combinations, see below or p. 103 of ref.[3]. On the other hand, any single determinant (open (= exists $\phi_i$ occupied by only one electron) or closed (= all $\phi_i$ occupied by two electrons) shell), particularly, $Y_k$, is always eigenfunction of $S_{op,z}$ [3], and $S_{op,z} Y_k = ((N_\alpha - N_\beta)/2)Y_k = M_S Y_k$.

**Ground state (K=k=0) LCAO coefficients a=1 vs. a=0 (TNRS)** are exhibited next via HF-SCF/STO-3G/a:
For $Ar^{6+}$: N=12, there are N/2=6 occupied MO i.e. i=1-6 in Eq.2, i=6[th] is the HOMO and i=7[th] is the LUMO.
For $Ar^{8+}$: N=10, there are N/2=5 occupied MO i.e. i=1-5 in Eq.2, i=5[th] is the HOMO and i=6[th] is the LUMO.
A little inconvenience has come up in the counting for $(\phi_i, \varepsilon_i)$: i=0,1,2,… in description e.g. in Eqs.1-6, 9-12, 16-22 or Tables 1-2), but in programming the i= 1,2,… is in use e.g. in LCAO coeff. matrices just below or Eqs.7-8, 13-15.
For $Ar^{6+}$, if a=1 ⇒ 4 convergence steps needed, $E_{electr,0}$= -509.818584 hartree, and the LCAO coefficients $\{c_{ij}\}$ in $S_0$ are

```
  εᵢ     -122.14152  -16.62156  -13.87757  -13.87757  -13.87757  -4.26796  -3.30475  -3.30475  -3.30475
j\i           1          2          3          4          5          6         7         8         9
1 AR 1S    -0.99482   -0.38605    0          0          0         -0.09245   0         0         0
2 AR 2S    -0.01453    1.05911    0          0          0          0.30052   0         0         0
3 AR 2PX    0          0          0.99287    0          0          0         0.28057   0         0
4 AR 2PY    0          0          0          0.12370    0.98513    0         0         0.16206   0.22903
5 AR 2PZ    0          0          0          0.98513   -0.12370    0         0        -0.22903   0.16206
6 AR 3S     0.00140    0.03268    0          0          0         -1.03031   0         0         0
7 AR 3PX    0          0          0.02753    0          0          0        -1.03138   0         0
```

```
8 AR 3PY  0         0         0         0.00343   0.02732   0         0        -0.59572  -0.84194
9 AR 3PZ  0         0         0         0.02732  -0.00343   0         0         0.84194  -0.59572
```
For $Ar^{8+}$, if a=1 ⇒ 4 convergence steps needed, $E_{electr,0}$= -500.696356 hartree, and the LCAO coefficients $\{c_{ij}\}$ in $S_0$ are
```
   ε_i    -123.74076 -18.11220 -15.39417 -15.39417 -15.39417 -4.85427  -4.34144  -4.34144  -4.34144
j\i            1         2         3         4         5         6         7         8         9
1 AR 1S    -0.99484   0.38610   0         0         0         0.09204   0         0         0
2 AR 2S    -0.01448  -1.05941   0         0         0        -0.29944   0         0         0
3 AR 2PX    0         0         0.27521   0.55842   0.77465   0        -0.04341   0.11532   0.24831
4 AR 2PY    0         0         0.93432   0.00920  -0.33856   0        -0.27119  -0.05268  -0.02295
5 AR 2PZ    0         0        -0.19741   0.82203  -0.52245   0        -0.03764   0.24651  -0.12107
6 AR 3S     0.00139  -0.03164   0         0         0         1.03034   0         0         0
7 AR 3PX    0         0         0.00666   0.01351   0.01874   0         0.16154  -0.42911  -0.92395
8 AR 3PY    0         0         0.02261   0.00022  -0.00819   0         1.00906   0.19601   0.08538
9 AR 3PZ    0         0        -0.00478   0.01989  -0.01264   0         0.14006  -0.91726   0.45049
```
For $Ar^{6+}$ if a=0 ⇒ <u>always</u> 1 convergence step needed, $e_{electr,0}$= -651.553609 hartree, and
for $Ar^{8+}$ if a=0 ⇒ <u>always</u> 1 convergence step needed, $e_{electr,0}$= -626.807313 hartree, and the LCAO coefficients in $Y_0$ are the same (the up-right 10x10 part of 12x12 determinant) for $Ar^{6+}$ and $Ar^{8+}$ (the effect of a=0):
```
   ε_i    -160.17164 -39.43488 -37.93238 -37.93238 -37.93238 -12.37315 -11.85348 -11.85348 -11.85348
j\i            1         2         3         4         5         6         7         8         9
1 AR 1S    -1.00273  -0.37183   0         0         0        -0.05947   0         0         0
2 AR 2S     0.00790   1.07962   0         0         0         0.21583   0         0         0
3 AR 2PX    0         0        -0.00223   0.88181  -0.50046   0        -0.12065  -0.12498   0.07921
4 AR 2PY    0         0        -0.89049   0.23761   0.42264   0         0.13731  -0.05649   0.12002
5 AR 2PZ    0         0         0.48485   0.44046   0.77393   0         0.05513  -0.13281  -0.12558
6 AR 3S    -0.00181  -0.04890   0         0         0        -1.02967   0         0         0
7 AR 3PX    0         0         0.00014  -0.05613   0.03186   0         0.65073   0.67407  -0.42721
8 AR 3PY    0         0         0.05668  -0.01513  -0.02690   0        -0.74059   0.30468  -0.64733
9 AR 3PZ    0         0        -0.03086  -0.02804  -0.04926   0        -0.29735   0.71633   0.67734
```
The $\varepsilon_i$/h of Eq.2 are also listed and Eqs.5-6 show how $S_0$ and $Y_0$ build up. MOs are occupied pair-wised with opposite spins by the N=10 or 12 electrons in ground state. The case a=0 is always obtained in one step, while a≠0 always needs many convergence steps increasing with size (N), (so e.g. one can alternatively start with a=0 to obtain $Y_0$ first and switch to a=1 and finishing the convergence for a=1 and $S_0$). Changing N, e.g. for $Ar^{8+}$ (N=10) to $Ar^{6+}$ (N=12), the $\phi_i$ (approximated via LCAO coefficients) and $\varepsilon_i$ change if a≠0, but remains strictly the same if a=0. If a=1 (generally a≠0) the LCAO coefficients matrix $[c_{ij}]$ in Eq.7 or list above forming columns for $\{\phi_i\}$ depend on N because $r_{ij}$ couples $(\varepsilon_i,\phi_i)$ and $(\varepsilon_j,\phi_j)$. If a=0, the columns are independent, in the example above, same for $Ar^{17+}$ (N=1), …$Ar^{8+}$, $Ar^{7+}$, $Ar^{6+}$ (N=12),…Ar (N=18) and up. The large difference in energies comes from including or omitting $H_{ee}$ ($e_{electr,0}$ << $E_{electr,0}$, "much larger" refers to CA). Comparing a=1 to a=0, irrelevant sign change (phase factor) happens, e.g. sign(-1.05941) vs. sign(1.07962) at j=2=i. Important is that TNRS can indicate the bond. The similar sets of LCAO coefficients in the two lists manifest. From Eqs.1-2, e.g. the closed shell $\rho(\mathbf{r}_1,a) \approx \rho(\mathbf{r}_1,\text{HF-SCF/basis}/a) = 2\Sigma_{i=1 \text{ to } N/2} \phi_i^2(\mathbf{r}_1,a)$ density does not vary strongly with "a", most importantly between a=0 and a=1, because of their close LCAO coefficients. The $\phi_i$'s are expanded in LCAO, and the enforced normalization in the algorithm, $\int\rho(\mathbf{r}_1,a)d\mathbf{r}_1$=N for any "a", makes the change via "a" even less visible. That is, the shape of $\rho(\mathbf{r}_1,a)$ does not change drastically with "a"; its integral properties change even less. For example, the electron-electron repulsion energy approx. right in Eq.18 in DFT or its alternative in HF-SCF formalism left in Eq.18 with J and K integrals does not change drastically either, LCAO phase factors drop by squares, a property important in Eq.9 below. However, the t(HF-SCF/basis/a)= $-\Sigma_{i=1}^{N/2}$ $<\phi_i(\mathbf{r}_1,a)|\nabla_1^2|\phi_i(\mathbf{r}_1,a)>$ kinetic energy can yield a more pronounced difference between a=0 and a=1, because the slopes $(\nabla_1\phi_i)$ differ (recall $<\phi_i|\nabla_1^2|\phi_i> = <\nabla_1\phi_i|\nabla_1|\phi_i>$). The electronic energy builds up quasi-linearly with simple curvature between $(Y_0,e_{electr,0})$ and $(\Psi_0,E_{electr,0})$ as a=0→1.

**The RHF/UHF mode in HF-SCF/basis/a=0 (TNRS):** For a=1, it is the most common molecular orbital method in practice for open shell molecules where the numbers of electrons in two spins are not equal. In case of triplet carbon atom ($1s^22s^22p_x^12p_y^1$, 2S+1=3) the UHF mode HF-SCF/STO-3G/a=1 MO energies

4 occupied MO with α spin: -10.93172  -0.72837  -0.32803  -0.32803   0.23600  /h
2 occupied MO with β spin: -10.88869  -0.46668   0.30854   0.38107   0.38107  /h

$E_{electr,0}$(HF-SCF): -37.198393 h, for theoretical S(S+1) = 2, a value of 2.0000 is obtained at STO-3G, but spin-contamination is 2.0048 at 6-31G**. The UHF mode HF-SCF/STO-3G/a=0 TNRS MO energies are

4 occupied MO with α spin: -17.76299  -3.92916  -3.67049  -3.67049  -3.67049  /h
2 occupied MO with β spin: -17.76299  -3.92916  -3.67049  -3.67049  -3.67049  /h

$e_{electr,0}$(HF-SCF): -50.725273 h, for theoretical S(S+1) = 2, a value of 2.0000 is also obtained at STO-3G and 6-31G**, i.e., no spin-contamination even at larger basis level. The UHF and RHF mode are the same in the case of HF-SCF/STO-3G/a=0, i.e. in the case of TNRS, that is, the coupled Roothaan equations, known as the Pople–Nesbet–Berthier equations fall back to simple Roothaan equations. This is, because the electron-electron repulsion causes the spatial part of MO to split in UHF (e.g. -10.93172 and -10.88869 above, etc.) to get deeper energy via VP in the case of a=1, more generally in the case of a≠0. The UHF method for a=1 (more generally for a≠0) mode in principle has this drawback.

The UHF virtually contradicts to that in $x_i$ and $x_j$ of spin-orbitals, for example, the spin coordinates $s_i$ and $s_j$ are enough to differ to satisfy the Pauli's exclusion principle. In the above HF-SCF/basis/a=1 numerical example, the spatial parts of MOs were allowed to split a bit to reach deeper energy. However, many scientists oppose to this split, such as UHF calculations based on that the $S_0$ single determinant, having an even worse form theoretically in the UHF mode than in the RHF mode. The profit on deeper energy via UHF vs. RHF mode in the HF-SCF/basis/a=1 calculation can be counterbalanced in DFT which applies the $E_{xc}[\rho(HF\text{-}SCF/basis/a=1)]$ functional during or after the algorithm, on the price of non-variational nature. However, the necessity of RHF/UHF mode trick annihilates as a→0.

**Convergence in HF-SCF/basis/a, a=0 vs. a≠0**: The HF-SCF/basis/a means that an SCF subroutine [2-3] performs the minimizations for H(a), particularly for $<S_0|H(a=1)|S_0>$ or $<Y_0|H(a=0)|Y_0>$ with single determinants keeping MO's orthonormal in it. However, $H(a=0)= H_\nabla+H_{ne}= \Sigma_{i=1}^N h_i$, so <u>SCF convergence is not needed at all</u>, as mentioned above, HF-SCF/basis/a=0 reduces to one convergence step to eigensolving Eq.2 for i+1=1,2,…K= N/2 or (N+1)/2, and not only $Y_0$, but excited $Y_k$ single determinants can be set up using the set $\{\phi_i\}$, e.g. if N=2, then

$$Y_0= |(\alpha\phi_0)(r_1),(\beta\phi_0)(r_2)>\text{ with } e_{electr,0}=2\varepsilon_0, \text{ and } S=2(\tfrac{1}{2} - \tfrac{1}{2} )+1=1, \quad (5)$$
$$Y_1= |(\alpha\phi_0)(r_1),(\alpha\phi_1)(r_2)>\text{ with } e_{electr,1}=\varepsilon_0+\varepsilon_1 \text{ and } S=3, \quad (6)$$

etc.. Importantly, $Y_k$ are all (k=0,1,…) eigenfunctions of Eq.1 with a=0, on the other hand, the same procedure for $S_k$ can be done, but nor $S_0$ nor $S_k$ are egenfunctions of Eq.1 with a=1 (generally with a≠0); more $S_k$ are worsening in quality as k>0. The electronic energy (see Eqs.5-6) is $e_{electr,k}= \Sigma_i n_i\varepsilon_i$, where $n_i$ is the population (0, 1 or 2), and $\Sigma_i n_i=$ N for ground and excited states. In contrast, $E_{electr,0}$(HF-SCF or KS/basis/a=1)≠ $\Sigma_i n_i\varepsilon_i$ for deepest possible filling in the single Slater determinant $S_0$, some cross terms must be subtracted [1-3].

In calculating $Y_k$, one has to apply standard linear algebra for (symmetric) energy Hamiltonian which requires to compute the matrix elements for (real) energy eigenvalues ($\varepsilon_i$) and orthonormal eigenvectors (MOs)

$$<b_i|h_1|b_j> \text{ for } i,j=1,2,…,K_b, \quad \text{and} \quad \phi_i(r_1)= \Sigma_{j=1}^{Kb} c_{ij}b_j(r_1), \quad (7)$$

where $\{b_1(r_1), b_2(r_1),…b_{Kb}(r_1)\}$ is an adequate, atom-centered AO basis set, and the set of LCAO coefficients is $\{c_{ij}\}$. The $K_b$ eigenvalues (MO energies) and eigenvectors (wave functions) of this square $K_b \times K_b$ Hamiltonian matrix approximates the lowest lying $K_b$ eigenstates via eigensolver: the orbital energy values, $\{\varepsilon_i\}_{i=1..Kb}$, and orthonormal (orbital or MO) wave functions, $\{\phi_i(r_1)\}_{i=1..Kb}$, (the numbering compatible with Eq.1 by i:=i-1). This is what we call a one-step algorithm, because the eigensolver is used only once. Now a=0, but in HF-SCF/basis/a≠0 algorithm every step after the initial estimation needs eigensolver, what HF-SCF do during a typical (a=1) SCF device (rotating the MO vectors with enforced orthonormality). Finally, the $\phi_i(r_1)$ wave functions (MOs) are expressed in LCAO in the basis set chosen, and are orthonormal. The eigenstate energy values $\{e_{electr,k}\}_{k=0,1,..L-1}$ along with the set of single Slater determinant wave functions $\{Y_k\}_{k=0,1,..L-1}$ can be accomplished systematically by mixing $\phi_i$ (i:=i+1= 1,2,…K_b), see Eqs.5-6, as in the standard algebra with a Slater determinant for case a=1, i.e. for $\{S_k\}$. All $\phi_i$ can be non-, singly- or doubly (oppositely) occupied by α or β spins and $K_b$>N, so L=(2K_b)!/(N!(2K_b-N)!) becomes huge as $K_b$ increases.

Statement: <u>In case of a=0, the N-electron system is accomplished in $Y_k$</u>, but the 3N spatial dimension $(r_1,…,r_N)$, in fact <u>4N spin-orbit dimension</u> $(x_1,…,x_N)$, partial diff. equation $(H_\nabla+ H_{ne})Y_k(x_1,…,x_N)=e_{electr,k}Y_k(x_1,…,x_N)$ in Eq.1 is <u>equivalent</u> with the <u>one-electron (N=1) 3 spatial dimension</u> $(r_1)$ partial diff. equation $h_1\phi_i(r_1)=\varepsilon_i\phi_i(r_1)$ in Eq.2 by the missing $r_{ij}$ terms via=0, examples of anti-symmetrized links between right of Eq.1 and Eq.2 are in Eqs.5-6. The Hamiltonian matrix is symmetric, that is $<b_i|h_1|b_j>= <b_j|h_1|b_i>$ for basis set elements ⇒ set of MO energies, $\{\varepsilon_i\}$, is a real valued set. Using Kronecker delta, the orthogonal properties are $<\phi_i|\phi_j>= <Y_i|Y_j>= \delta_{ij}$, as well as

$$\varepsilon_j\delta_{ij}= <\phi_i|h_1|\phi_j>= <\phi_j|h_1|\phi_i>= \varepsilon_i\delta_{ji} \quad \text{and} \quad e_{electr,j}\delta_{ij}= <Y_i|H_\nabla+H_{ne}|Y_j>= <Y_j|H_\nabla+H_{ne}|Y_i>= e_{electr,i}\delta_{ji}. \quad (8)$$

The evaluation of the symmetric Hamiltonian matrix elements $<b_i|h_1|b_j>$ can be done with Gaussian (GTO: $p(r_1)\exp(-a|R_A-r_1|^2)$) or Slater (STO: $p(r_1)\exp(-a|R_A-r_1|)$) basis sets $\{b_i(r_1)\}$, i.e. with atom centered ones, (p is polynomial). Use of GTOs allows analytical integration, while STOs need numerical integration, both yielding numerical (particularly LCAO) approx. by eigensolving Eq.2 for $\phi_i$, M≥0. Hydrogen-like atoms (M=1 in Eq.2) have STO analytical solutions for ground (i=0) and excited states (i>0): the normalized $\phi_0=p(r_1)E(r_1)$ with $E=\exp(-Z_AR_{A1})$, $p= Z_A^{3/2}/\pi^{1/2}$ and $\varepsilon_1=-Z_A^2$ called 1s, or $\phi_1=p(r_1)E(r_1)$ with $E=\exp(-Z_AR_{A1}/2)$, $p= (2-Z_AR_{A1})Z_A^{3/2}/(32\pi)^{1/2}$ and $\varepsilon_1=-Z_A^2/4$ called 2s orbitals, etc.. By this reason, STOs provide faster convergence; in practice linear combination of e.g. 3-6 GTOs is fitted to an STO exponential in integral sense. Unfortunately analytical solution for M>1 in Eq.2 is not known yet (nor for i=0, nor for i>0), but if it was known, one could avoid numerically solving Eq.2 over and over again for the right of Eq.1 by e.g.

HF-SCF/basis/a=0 algorithm for an LCAO approx. The M=2 case of Eq.2, the hydrogen molecule ion ($H_2^+$) was targeted heavily early in the history of computation chemistry in many ways, but compact analytical expression was not found. However, for example, for M>1, the not atom centered $\phi_i = p(\mathbf{r}_1)E(\mathbf{r}_1)$ with $E=\exp(-\Sigma_{A=1}^M Z_A R_{A1}/n_A)$ can be a powerful candidate for semi-analytic solution (with $n_A$ not necessarily integer), where $p(\mathbf{r}_1)$ is a polynomial to fit via the Hamiltonian matrix of Eq.2: Recall that e.g. for all $n_A=1$ for ground state, the identity $\nabla_1^2(pE) = (\nabla_1^2 p)E + 2\nabla_1 p \nabla_1 E + p(\nabla_1^2 E)$ yields $p(\nabla_1^2 E) = -pE[(\Sigma_A Z_A(x_1-R_{Ax})/R_{A1})^2 + (.)^2 + (.)^2] - 2pE(\Sigma_A Z_A R_{A1}^{-1})$ in which the 2$^{nd}$ term with proper coefficient can cancel $H_{ne}(N=1) = \Sigma_A Z_A R_{A1}^{-1}$ in Eq.2 just as in the case of H-like atoms.

## CALCULATING GROUND STATE FOR a=1 WITH THE HELP OF a=0

Manipulating with $<\Psi_K|H_{ee}|Y_k> = <\Psi_K|H(a=1) - (H_\nabla + H_{ne})|Y_k>$ one obtains for K and k excited states that $E_{electr,K} = e_{electr,k} + (N(N-1)/2)<\Psi_K|r_{12}^{-1}|Y_k>/<\Psi_K|Y_k>$, and using VP for K=k=0:

$$E_{electr,0} \approx E_{electr,0}(TNRS) \equiv e_{electr,0} + <Y_0|H_{ee}|Y_0> = e_{electr,0} + (N(N-1)/2)<Y_0|r_{12}^{-1}|Y_0> . \qquad (9)$$

Eq.9 is the simplest and immediate estimation for a=1 from TNRS (a=0), and can be improved for example with DFT, CI, etc. devices. Tests on Eq.9 have shown that, it brings the $e_{electr,0}$ very close to $E_{electr,0}$, but improvement is needed to reach chemical accuracy (CA= 1 kcal/mol= 1.593 mili h). In the vicinity of stationary points $E_{electr,0} - e_{electr,0} \approx N(N-1)/5$ h, i.e. huge, but Eq.9 reduce it to $N(N-1)/50$ h, i.e. by a magnitude.

**Generalization of Møller-Plesset (MP) perturbation theory (1934)** [3] from HF-SCF/basis/a=1 can be made for any "a". This theory has been proved originally for a=1, we state here that it applies for any "a", as well as the a=0 is a special case. The essential observation in MP perturbation theory (a≠0) is that, all Slater determinants formed by exciting electrons forming the occupied to the virtual orbitals are also eigenfunctions of VNRS (see discussion with Eq.2) with an eigenvalue equal to the sum of the one electron energies of the occupied spin-orbitals, so a determinant formed by exciting from the p$^{th}$ spin-orbital in the Hartree-Fock ground state into the r$^{th}$ virtual spin-orbital only changes the MO in the Slater determinant, and the eigenvalue changes to $E_{electr,0}$(HF-SCF/basis/a)$\to E_{electr,0}$(HF-SCF/basis/a)$+ \varepsilon_p - \varepsilon_r$, which is somewhat surprising but true for any "a". Especially important if a=1, and trivial if a=0. The typical corrections in MP theory take the form: $|<s_0(a)|ar_{ij}^{-1}|s_{0,pq}^{rs}(a)>|^2/(\varepsilon_p + \varepsilon_q - \varepsilon_r - \varepsilon_s)$ in the second order correction, wherein the $s_{0,pq}^{rs}(a)$ is the determinant from $s_0(a)$ e.g., the p and q spin-orbitals are changed to the excited (virtual) r and s ones, as well as this term is summed up for all i<j electrons (N) and all r<s available (i.e. calculated) virtual spin-orbitals. This expression is extended with the coupling strength parameter "a" and particularly for a=0 all the MP corrections cancel (because of the term $ar_{ij}$), which means that a=0 does not need any correction, because the Slater or single determinant is an accurate wave function form. However, for practical use, it needs manipulation to relate the case a=0 to a=1 somehow. For example, Eq.9 includes the electron-electron repulsion in simplest form for a main term, but it must be refined further to account better for the correlation effect (recall CA). MP theory, for example, can provide refinement over Eq.9 as follows.

In relation to the MP theory, the $s_0(a)$ which includes the effect of $ar_{12}^{-1}$ somehow (a≠0), but not precisely, the MP tries to correct it to approach $y_0(a)$ most importantly $\Psi_0(a=1)$ as possible in terms of energy; however, $Y_0(a=0)$ is not affected by $r_{12}^{-1}$ at all. MP theory corrects what HF-SCF/basis/a≠0 makes in approximating electron-electron repulsion energy, i.e. the error coming from $<\Psi_0|H_{ee}|\Psi_0> \approx <S_0|H_{ee}|S_0>$ when e.g., a=1, based on the Rayleigh-Schrödinger perturbation theory. The case a=1 is generalized here for general a≠0. One point is fundamental in applying MP for a=0 case: The first order MP energy correction [3] is $E_0^{(1)} = <s_0(a\neq0)|aH_{ee}-aV_{aa,eff}|s_0(a\neq0)>$, i.e., only the difference is in the core (see $aV_{aa,eff}$ in discussion with Eq.2). If a=0, the $E_0^{(1)}=0$, i.e. a=0 case needs no correction but, if we want to relate a=0 to a=1, then "$aH_{ee}$ is not approximated by $aV_{aa,eff}$", but "$H_{ee}$ is approximated crudely with zero", because a=0, so $E_0^{(1)} = <Y_0(a=0)|H_{ee}-0|Y_0(a=0)>$, i.e., the full term is in the core, exactly what Eq.9 has from another point of view. Correction to Eq.9 with higher accuracy can be done with the exact MP2 analogue $|<Y_0|r_{12}^{-1}|Y_{0,pq}^{rs}>|^2/(\varepsilon_p+\varepsilon_q-\varepsilon_r-\varepsilon_s)$ terms, as well as with MP3, MP4, etc. terms, if MP method is chosen for correction.

Known in HF-SCF/basis/a=1 that the 1$^{st}$ order MP is only the HF-SCF energy itself, an overlap between HF-SCF and MP theory. Similarly in TNRS, the 1$^{st}$ order perturbation to a=0 to approximate a=1 for K=0 ground state is just Eq.9, this latter can also be deduced simply if the $Y_0$ as trial function is substituted for $\Psi_0$ into $E_{electr,0} = <\Psi_0|H(a=1)|\Psi_0>$. The one-electron density from a=0 to a=1 can also be perturbed as the electronic energy for ground state discussed: Adding weighted terms as $\rho_0(\mathbf{r}_1,a=1) \approx \rho_0(\mathbf{r}_1,a=0) + \Sigma C_{pqrs} \int Y_0^* Y_{0,pq}^{rs} ds_1 d\mathbf{x}_2 \ldots d\mathbf{x}_N$ to estimate on "MP2 level" looks plausible. The integration gives the expected $\int \rho_0(\mathbf{r}_1,a=0)d\mathbf{r}_1 + \Sigma C_{pqrs}<Y_0|Y_{0,pq}^{rs}> = N + \Sigma 0 = N$, since $\{Y_k\}$ is an orthonormal set. In this way, the estimation does not have to be re-normalized, its shape is corrected by "add-subtract" design which keeps the norm, however, its derivative changes, because $\nabla_1[\rho_0(\mathbf{r}_1,a=0) + \Sigma C_{pqrs} \int Y_0^* Y_{0,pq}^{rs} ds_1 d\mathbf{x}_2 \ldots d\mathbf{x}_N] \neq \nabla_1 \rho_0(\mathbf{r}_1,a=0)$, necessary for better kinetic energy estimation, (the latter increases as "a" decreases).

**Utilizing square of Hamiltonian operator for ground state** (K=0) is possible directly for a=1 and for pre-calculation a=0 to estimate a=1. Applying the Hamiltonian twice for the ground state (and a=1) wave function simply yields $H^2\Psi_0 = E_{0,electr}H\Psi_0 = E_{0,electr}{}^2\Psi_0$, that is $E_{0,electr} = (\langle\Psi_0|H^2|\Psi_0\rangle)^{1/2}$. The $H^2$ preserves the linearity and hermetic property from operator H, and if e.g. HF-SCF single determinant $S_0$ approximates $\Psi_0$ via variation principle from $\langle S_0|H|S_0\rangle$, the approx. $E_{0,electr} \approx (\langle S_0|H^2|S_0\rangle)^{1/2}$ is better than $E_{0,electr} \approx \langle S_0|H|S_0\rangle$, and similarly, $E_{0,electr} \approx \langle Y_0|H^2|Y_0\rangle^{1/2}$ is better than $E_{0,electr} \approx \langle Y_0|H|Y_0\rangle$, (by VP the $\langle S_0|H|S_0\rangle < \langle Y_0|H|Y_0\rangle$), coming from basic linear algebraic properties of linear operators for the ground state. The $H(a)^2 = (H_\nabla + H_{ne} + aH_{ee})^2$ is hectic by the non-associative $H_\nabla H_{ee}$ and $H_\nabla H_{ne}$, so using $\langle y_0|H(a)^2 y_0\rangle = \langle H(a)^2 y_0|y_0\rangle = \langle H(a)y_0|H(a)y_0\rangle$, the far right side keeps the algorithm away from operators like $\nabla_1{}^2 r_{12}{}^{-1}$, etc. at least. With grouping $(H_\nabla + H_{ne}) + aH_{ee} = \Sigma_{i=1}{}^N h_i + aH_{ee}$, the $\langle H(a)y_0|H(a)y_0\rangle = \langle\Sigma_{i=1}{}^N h_i\, y_0|\Sigma_{i=1}{}^N h_i\, y_0\rangle + 2a\langle\Sigma_{i=1}{}^N h_i\, y_0|H_{ee}y_0\rangle + a^2\langle y_0|H_{ee}{}^2 y_0\rangle$. The Coulomb integrals with operators $H_{ne}{}^2$, $H_{ne}H_{ee}$ and $H_{ee}{}^2$ come up, that is with distance operators $R_{Ai}R_{Bj}$, $R_{Ai}r_{jk}$ and $r_{ij}r_{kl}$ with A,B=1..M and i,j,k,l=1…N, all can be evaluated numerically with STO basis sets [4] or analytically with GTO basis sets [3-4], Appendix details $\langle y_0|H_{ee}{}^2|y_0\rangle$.

For $y_0 \approx S_0$ from HF-SCF/basis/a=1 the integrals for $\langle S_0|H(a=1)^2|S_0\rangle$ can be evaluated, but no further convenient reduction. However, for $y_0 = Y_0$ from HF-SCF/basis/a=0, the formulas reduce further via Eq.1 and $\langle Y_0|Y_0\rangle = 1$ as

$$E_{electr,0}{}^2 \approx [E_{electr,0}(\text{TRNS with } H^2 \text{ trick})]^2 \equiv \langle Y_0|H(a=1)^2|Y_0\rangle = e_{electr,0}{}^2 + 2 e_{electr,0}\langle Y_0|H_{ee}|Y_0\rangle + \langle Y_0|H_{ee}{}^2|Y_0\rangle . \quad (10)$$

With the very crude Eq.21 (in Appendix with $y_0 = Y_0$) Eq.10 reduces to Eq.9 on the price of losing the improvement by $H^2$ trick. The more accurate Eq.20 (in Appendix) yields larger absolute value than Eq.21 with adding two positive terms, providing deeper negative square root in Eq.10 than the value by Eq.9, see also Eq.22. (One must use negative sign for bound state when taking the square root in Eq.10 in estimating $E_{electr,0}$.) The idea with Eq.10 can be continued as $(E_{electr,0})^n \approx \langle Y_0|H(a=1)^n|Y_0\rangle$, but evaluation of $\langle Y_0|H_{ee}{}^n|Y_0\rangle$ becomes very difficult if n>2.

Empirical treatment/improvement of Eq.10 as well as decreasing the computation can be made with e.g. $\langle Y_0|H_{ee}{}^k|Y_0\rangle \approx c_k\langle Y_0|H_{ee}|Y_0\rangle^k$ (where $c_1 = 1$ with k=2 is the weak approx. Eq.21) if the value of empirical $c_k$ (k=2,…n, but at least for k=2) was quasi-independent of molecular frame ($H_{ne}$) and yielded good approx., (but one generally faces to the convergence problem known in "moment of one-electron density" DFT device). Eq.18 is a basic DFT approximating device and can be used in Eq.10. Using $z_0 \equiv \langle Y_0|H_{ee}|Y_0\rangle$, Eq.9 is just the 1st level moment approx. $e_{electr,0} + z_0$, Eq.10 reduces to 2nd level moment approx. $-(e_{electr,0}{}^2 + 2e_{electr,0}z_0 + c_2 z_0{}^{w2})^{1/2}$ which further reduces to Eq.9 if $c_2 = 1$ and $w_2 = 2$ but empirical values improve, and the 3rd level moment approx. is $-(e_{electr,0}{}^3 + 3e_{electr,0}{}^2 z_0 + 3c_2 e_{electr,0} z_0{}^{w2} + c_3 z_0{}^{w3})^{1/3}$ which also reduces to Eq.9 if $c_2 = c_3 = 1$, $w_2 = 2$ and $w_3 = 3$ but empirical values also improve. Empirical parameter fit for $c_k$ and $w_k$ was done using the set of 185 atomic ions ($1 \leq N, Z_A \leq 18$, M=1, $E_{electr,0}(CI) = -0.5$ to -527.544 hartree for H to Ar atoms, resp.), its statistics is listed in Table 1 to test the TNRS method using HF-SCF/6-31G*/a=0 for $Y_0$. For value 0.3951, the 2nd order (0.5171) improves over the 1st (1.7791), but the 3rd (0.6977) is not better than the 2nd. It is interesting that $c_2$ is close to 1 but necessary for accuracy and $w_2$ is very close to 2 by the fit, that is, $\langle Y_0|H_{ee}{}^2|Y_0\rangle \approx c_2\langle Y_0|H_{ee}|Y_0\rangle^2$ with $c_2 \approx 1$, as well as the $w_3$ is surprising: The fit "tends to avoid" value 3 (basic problem of moments with high power). As a consequence, one should use Eq.10 for increased accuracy on the price of a bit more difficult integral evaluation.

We need Eqs.17 and 19 (in Appendix) for $y_0(a=0) = Y_0$ and $y_0(a=1) \approx S_0$. Using Eq.17 in Eq.10, the VP still holds, but using Eq.19 in Eq.10 makes it non-variational. Grouping $H = H_\nabla + (H_{ne} + H_{ee})$ and using $\langle S_0|H(a=1)^2|S_0\rangle = \langle HS_0|HS_0\rangle$ to avoid $H_\nabla H_{ne}$ and $H_\nabla H_{ee}$, the analog expression to Eq.10, i.e. direct calculation with a=1 is

$$E_{electr,0}{}^2 \approx [E_{electr,0}(\text{HF-SCF/basis/a=1 with } H^2 \text{ trick})]^2 \equiv \langle S_0|H(a=1)^2|S_0\rangle =$$
$$\langle H_\nabla S_0|H_\nabla S_0\rangle + 2\langle H_\nabla S_0|(H_{ne}+H_{ee})S_0\rangle + \langle S_0|(H_{ne}+H_{ee})^2|S_0\rangle . \quad (11)$$

Also, the last term in Eq.11 can be evaluated analogously to Eq.17 and the VP holds, however, using Eq.19 makes it non-variational. The 1st and 2nd terms in Eq.11 are not as simple as the corresponding ones in Eq.10, but at least these can be evaluated in similar standard ways, ($H_\nabla$ on GTO preserves the analytical form of GTO).

**Configuration interactions (CI) from TRNS focusing on ground state**. The mathematical case a=0 generates an orthonormalized set of single Slater determinants $\{Y_k(a=0)\}$ which can be used as basis set for CI calculations for the physical case a=1 in Eq.1. The excited states $Y_k$ with k>0 can be obtained via algorithm HF-SCF/basis/a=0 [4] or directly solving Eq.2 by eigensolver and building $Y_k$ up along with Eqs.5-6. However, in both methods one has to use basis set $\{b_i\}$ adequate for higher $\phi_i$ (i > N/2 or (N+1)/2, i.e. LUMO+1, +2, etc.) states. One can simply increase N by e.g. 1 or 2 (or more for k>>0) for the same nuclear frame. For example in case of Ar above, calculation for Ar⁻ (N=19) must be performed (even it is not a stable ion is nature; the 19th electron is a "virtual" one) if one wants to improve HF-SCF/basis/a=1 calculation for neutral Ar (N=18) with CI based calculation: It means that the basis set should include 3d and 4s basis functions as well, (for $Y_0$ the minimal basis set is based on ground state Ar atom configuration $1s^2 2s^2 2p^6 3s^2 3p^6$). This $\{Y_k\}$ determinant basis set can be used for CI calculations as the $\{S_k(a=1)\}$ in practice, the linear algebra is exactly the same, but the algebraic forms do differ slightly (and simpler), see below.

The standard way of expanding anti-symmetric wave functions $\Psi_K$ (of a=1, i.e. left in Eq.1, most importantly to ground state K=0) using the orthonormal N-electron determinant basis set $\{Y_k\}$ (of a=0, i.e. right in Eq.1, or Eq.2 with Eqs.5-6, or from HF-SCF/basis/a=0) is the linear combination of single, double, triple, etc. excited N-electron Slater determinants: To the conventional $\Psi_K = \Sigma_k d_k(K) S_k$ with $\{S_k\}$ from HF-SCF/basis/a=1, the alternative is

$$\Psi_K = \Sigma_k c_k(K) Y_k, \quad \text{especially} \quad \Psi_0 = \Sigma_k c_k(0) Y_k, \quad \text{and} \quad k=0,1,2,\ldots L-1, L, \ldots \infty. \quad (12)$$

By the principles of linear algebra, changing basis set ($\{S_k\} \to \{Y_k\}$) should not be a problem, mainly from the point of slowly changing LCAO parameters in the range $0 \le a \le 1$. The routine generation [3] of single determinant basis set, e.g. for singlet (2S+1=1) excitation, is taking any 5 columns from the i=1,…,6 (allowing the LUMO to be used only) for an NxN= 10x10 determinant $Y_k$ (a=0) or $S_k$ (a=1) in the above lists of LCAO for $Ar^{8+}$ (N=10): For (a chosen particular level) CI, the ground state (k=0) determinant is $|(\alpha\phi_1)(r_1), (\beta\phi_1)(r_2), \ldots, (\alpha\phi_5)(r_9), (\beta\phi_5)(r_{10})\rangle$ with $\phi_i(r_1) = \Sigma_{j=1}^{9} c_{ij}(STO-3G)_j(r_1)$, the LCAO coefficients $\{c_{i1}, \ldots, c_{i9}\}$ for $S_0$ or $Y_0$ from the list above, and see Table 2 for some other configurations for N=10 electrons using the lowest lying 6 levels (i-1=0,1,…,5) with double excitation singlets.

Standard linear algebra approximates the set of eigenstates $\{\Psi_K, E_{electr,K}\}$ for a=1 by expanding $\Psi_K$ in basis set $\{Y_k\}$: The first main step is the diagonalization/eigensolving of Hamiltonian matrix $\langle b_i|h_1|b_j\rangle$ in Eq.7 and to set up eigenstates $\{Y_k, e_{electr,k}\}$ as single determinant basis set as in Eqs.5-6. (The algorithm HF-SCF/basis/a=0, with tricking (virtual) N, can be used for this first step also, although its SCF part will not be called.) The second main step is the diagonalization/eigensolving of Hamiltonian matrix with elements (using Eq.8)

$$\langle Y_{k'}|H_\nabla + H_{ne} + H_{ee}|Y_k\rangle = e_{electr,k} \langle Y_{k'}|Y_k\rangle + \langle Y_{k'}|H_{ee}|Y_k\rangle. \quad (13)$$

The diagonal elements (k'=k) reduce to the generalization of Eq.9 for ground (K=k=0) and excited (K,k>0) states

$$E_{electr,K} \approx E_{electr,K}(TNRS) \equiv \langle Y_k|H_\nabla + H_{ne} + H_{ee}|Y_k\rangle = e_{electr,k} + (N(N-1)/2)\langle Y_k|r_{12}^{-1}|Y_k\rangle. \quad (14)$$

The off-diagonal elements (k'≠k) show purely Coulomb electron-electron interaction terms

$$\langle Y_{k'}|H_\nabla + H_{ne} + H_{ee}|Y_k\rangle = e_{electr,\, k\, or\, k'}\langle Y_{k'}|Y_k\rangle + \langle Y_{k'}|H_{ee}|Y_k\rangle = (N(N-1)/2)\langle Y_{k'}|r_{12}^{-1}|Y_k\rangle. \quad (15)$$

Eq.9 is the (1,1) element of the CI matrix (of any level), as another verification of Eq.9, above $Y_0$ was trial function for $\Psi_0$. We will not review and discuss the different levels (full-, Multi-Reference Single and Double (MRSD)-, etc.) of CI [3] calculations, only a simple one showing the idea related to TNRS. In today practice, where not the $\{Y_k\}$ by HF-SCF/basis/a=0, but $\{S_k\}$ by HF-SCF/basis/a=1 is used, the off-diagonal elements corresponding to Eq.15 contain orbital energies $\varepsilon_i$ of MOs too. The CI matrix in Eq.13 is diagonal for the right one (a=0) in Eq.1, because the set of wave functions $\{Y_k\}$ is expressed with itself, as triviality. Neglecting off-diagonal elements (Eq.15), the matrix in Eq.13 diagonalizes to approx. Eq.14, but it is below CA. To reach CA, the full matrix must be used in Eq.13 for correcting basis set error and correlation effect in Eq.14 via the off-diagonal elements (or cross terms), as well as with a larger size matrix in Eq.13, not only the ground state but (lower lying) excited states can also be estimated. Restriction by Brillouin theorem applies only if a≠0, however $\{Y_k\}$ originates from a=0, that is, single excited determinants in $Y_k$ can already be used (4a and 4b in Table 2) in the expansion, not only doubly excited and up ones. The $\langle Y_{k'}|r_{12}^{-1}|Y_k\rangle$ terms generate many products, but the orthogonality of MOs in $\{Y_k\}$ makes many cancellations [3]; the spin related properties and manipulations [3] are exactly the same in both, $\{Y_k\}$ and $\{S_k\}$. The ground state one-electron density (using notation $\int_{(1)} (.) \equiv \int (.) ds_1 dx_2 \ldots dx_N$) is $\rho_0(r_1, a=0) = N\int_{(1)} Y_0^* Y_0$ and

$$\rho_0(r_1, a=1) = N\int_{(1)} \Psi_0^* \Psi_0 \approx N\int_{(1)} [\Sigma_k c_k(0) Y_k]^2 = N\int_{(1)} \Sigma_k [c_k(0) Y_k]^2, \quad (16)$$

wherein the orthonormality (Eq.8) is used and the approx. is good if k goes up to high enough value. The first crude approx. of $\rho_0(r_1, a=1)$ is just the $\rho_0(r_1, a=0)$, like the Eq.9 for $E_{electr,0}$.

Eq.14 contains only one single determinant, so its spin state is obvious, however for any kind of CI correction the spin situation must be taken into account. Eq.1 does not contain spin coordinates, but Eqs.3-4 hold, and must be taken into account when using Eq.13 for calculation, see also the discussion above with Table 2. As usual, e.g. for a singlet state (2S+1=1) molecule, those $Y_k$ determinants must be eliminated from the determinant expansion which are not singlets ($M_S \neq 0$). The spin algebra [3] is the same for Slater determinants $\{Y_k\}$ and $\{S_k\}$. For example, the two simplest spin-adapted cases for even N in $Y_0$ obtained from HF-SCF/basis/a=0: the doubly excited singlet $Y_{p(\alpha)p(\beta)}^{r(\alpha)r(\beta)}$, wherein $(\alpha,\beta)$ electron pair from p orbital below LUMO are promoted to r orbital over HOMO with the same $(\alpha,\beta)$ spin configuration as indicated in brackets (e.g. N=10, k=0-3 in Table 2), and the singly excited singlet configuration is $(Y_{p(\alpha)}^{r(\alpha)} + Y_{p(\beta)}^{r(\beta)})/2^{1/2}$, in the latter the two terms alone are also diagonal elements, but not pure spin states, (e.g. k=4a and 4b in Table 2).

Table 2 shows two examples, N=9 and N=10 electron systems, for which only the LUMO state is used for CI. Only some of those energetically lowest lying configurations are shown for excited determinants $Y_k$, which have same spin state (2S+1) to use in Eq.12 for $\Psi_0$. If $k \to \infty$ in Eq.12, the equality holds, if one stops at a reasonable L, that is a good approx. only. Eq.9 is remarkably, but to reach the CA, one should use at least 2, 3 or 4 (L=1, 2 or 3) elements and at least the lowest lying excited state LUMO, i.e. set $\{Y_k\}_{k=0}^{1or2or3}$ from Table 2 in the expansion Eq.12, (for other N values, the top of population/configuration is the same). Recall that excited determinants ($Y_k$ with k>0, i.e. including at least a LUMO) necessary [5] for estimating ground state $\Psi_0$, (in contrast to DFT which expands with only using ground state one-electron density, $\rho_0(\mathbf{r}_1,\text{HF-SCF/basis}/a=1)$, for correlation effect in $S_0$). With L=1,2,3, Eqs.12-13 reduce to $2^{nd}$, $3^{rd}$ and $4^{th}$ order algebraic equations, one eigensolver for Eq.2 is enough, no need the other one for Eq.13. However, for L>3 one needs eigensolver for Eqs.12-13 to approximate ($E_{electr,K}$, $\Psi_K$) for K=0, 1 or 2, for example, (to treat higher than $4^{th}$ order algebraic equation). Eq.9 is the case L=0 in Eq.12, to correct its basis set error and correlation effect on L=1 first level (k=0,1) is as follows: Using Table 2 for (general N and) same spin states (same 2S+1 in all chosen $Y_k$, k=0,1…L) especially for HOMO→LUMO excitation, the symmetric 2x2 Hamiltonian is $H_{11}\equiv e_{electr,0}+<Y_0|H_{ee}|Y_0>$, $H_{12}\equiv <Y_0|H_{ee}|Y_1>=H_{21}$, $H_{22}\equiv e_{electr,0}-2\varepsilon_{HOMO}+2\varepsilon_{LUMO}+<Y_1|H_{ee}|Y_1>$, the $(H_{11}-\lambda)(H_{22}-\lambda)-H_{12}^2=0$ secular equation provides two real roots, the lower one is an estimation for $E_{electr,0}$ giving lower (VP) value than Eq.9. Because L=1 only, the higher root is only a weak estimation for next level excited states with same spin state (not necessarily $E_{electr,1}$, because other spin state with its lower value can be a candidate for $E_{electr,1}$, see e.g. Hund's rule).

Another solution for Eqs.12-13 for ground state (K=0) can come from the idea of quadratically convergent HF-SCF (or QC-SCF) orbital optimization: $E_{electr,0}\approx <\Sigma Y_k|H(a=1)|\Sigma Y_k> = <\Sigma Y_k|\Sigma(e_{electr,k}+H_{ee})Y_k> = \Sigma^2 e_{electr,k}+<\Sigma Y_k|H_{ee}|\Sigma Y_k>$ where sum operators are $\Sigma\equiv\Sigma_{k=0}^L c_k(0)$ and $\Sigma^2\equiv\Sigma_{k=0}^L [c_k(0)]^2$ and must be minimized w/r to $\{c_k(K=0)\}_{k=0}^L$, using spin states (2S+1) of the lowest possible filling for all chosen element in basis $\{Y_k\}$ by the VP, see also Table 2 for guiding; the L=0 yields Eq.9 again with $c_0(0)=1$. In this case the K=0 ground state is targeted only by choosing a reasonable low L value along with LUMO, LUMO+1, etc.. The condition is normalization $<\Sigma Y_k|\Sigma Y_k>=1$ for the minimum $\partial<\Sigma Y_k|H(a=1)|\Sigma Y_k>/\partial c_i(0)=0$ for i=0,1,…L, so for example the Lagrange multiplier method can be used. The orthonormality in Eq.8 reduces the normalization to the simpler $\Sigma^2 1=1$, and the Lagrangian is Lgr= $\Sigma^2 e_{electr,k}+<\Sigma Y_k|H_{ee}|\Sigma Y_k>+\lambda(1-\Sigma^2 1)$, finally, $\partial Lgr/\partial c_i(0)=0$ for i=0,1,…L and $\partial Lgr/\partial\lambda=1-\Sigma^2 1=0$, a $3^{rd}$ order equation system (by terms $\lambda[c_i(0)]^2$) for which standard subroutines are available (e.g. newton slope method, etc.).

As a final word in this part for CI, eigensolving matrix in Eq.7 provides the single determinant basis set $\{Y_k\}$ via Eqs.5-6, thereafter eigensolving matrix in Eq.13 buys off the needs of SCF convergence (rotation of orthonormal vectors of LCAO coefficients $[c_{i1},…c_{iKb}]^T$, i=1…$K_b$ in Eq.7) typically performed in the standard HF-SCF/basis/a=1 algorithm [1-3], as well as these two eigensolvers correct the basis set error and correlation effect present in the latter.

## APPENDIX

The cardinality of the set generated by electron-electron repulsion operator $H_{ee}^2$ comes from elementary combinatorics: $H_{ee}$ contains $\binom{N}{2}=N(N-1)/2$ and $H_{ee}^2$ contains $N^2(N-1)^2/4$ terms. For two different excited single determinants, e.g. $y_K(a=1)\approx S_{k=K}$ or $y_k(a=0)=Y_k$ from HF-SCF/basis/a=1 or 0, resp., generally for two different ground or excited anti-symmetric wave functions, the integration contains three kinds of terms: $r_{12}^{-2}$, $r_{12}^{-1}r_{13}^{-1}$ and $r_{12}^{-1}r_{34}^{-1}$ as

$$<y_K|H_{ee}^2|y_{K'}> = \binom{N}{2}\{<y_K|r_{12}^{-2}|y_{K'}> +2(N-2)<y_K|r_{12}^{-1}r_{13}^{-1}|y_{K'}> +\binom{N-2}{2}<y_K|r_{12}^{-1}r_{34}^{-1}|y_{K'}>\}. \quad (17)$$

The control sum $\binom{N}{2} +2(N-2)\binom{N}{2} + \binom{N}{2}\binom{N-2}{2} = N^2(N-1)^2/4$ holds, as well as the magnitude of cardinality of individual terms on the right in Eq.17 are $N^2$, $N^3$ and $N^4$, resp.; (the last term on the right breaks further as $\int$(seed $r_{12}^{-1}r_{34}^{-1}$)= $\int$(seed $r_{12}^{-1}$)$\int$(seed $r_{34}^{-1}$) as e.g. in Eq.19). Eq.17 reduces further in standard way [3] for final evaluation e.g. with GTO.

Focusing on ground state (K=K'=0), using notations $\int_j(.)\equiv \int(.)(\Pi_{i=1}^j d\mathbf{r}_i)$ and keeping in mind that $<y_0|y_0>=1$ and $\int_j \Pi_{i=1}^j \rho_0(\mathbf{r}_i,a)=N^j$ with j=1,2,3,4 $\Rightarrow$ $<y_0|H_{ee}|y_0> = \binom{N}{2}<y_0|r_{12}^{-1}|y_0> \approx \binom{N}{2}N^{-2}\int_2\rho_0(\mathbf{r}_1,a)\rho_0(\mathbf{r}_2,a)r_{12}^{-1} = ((N-1)/(2N))\int_2\rho_0(\mathbf{r}_1,a)\rho_0(\mathbf{r}_2,a)r_{12}^{-1} \approx (½)\int_2\rho_0(\mathbf{r}_1,a)\rho_0(\mathbf{r}_2,a)r_{12}^{-1}$ as

$$<y_0|H_{ee}|y_0>= (N(N-1)/2)<y_0|r_{12}^{-1}|y_0>\approx ½\int \rho_0(\mathbf{r}_1,a)\rho_0(\mathbf{r}_2,a)r_{12}^{-1}d\mathbf{r}_1 d\mathbf{r}_2. \quad (18)$$

Eq.18 is the emblematic Coulomb expression á la DFT especially for a=1, a form which has a vast literature, called exchange-correlation calculation to correct "$\approx$" to "=" [1-2]. In similar manner, the DFT approx. for the three terms in Eq.17 is as follows: The $1^{st}$ term has the same form as Eq.18 but with $r_{12}^{-1}\rightarrow r_{12}^{-2}$, the $2^{nd}$ is $2\binom{N}{2}(N-2)<y_0|r_{12}^{-1}r_{13}^{-1}|y_0>\approx 2\binom{N}{2}(N-2)N^{-3}\int_3(\Pi_{i=1}^3\rho_0(\mathbf{r}_i,a))r_{12}^{-1}r_{13}^{-1}= ((N-1)(N-2)/N^2) \int_3(\Pi_{i=1}^3\rho_0(\mathbf{r}_i,a))r_{12}^{-1}r_{13}^{-1}\approx \int_3(\Pi_{i=1}^3\rho_0(\mathbf{r}_i,a))r_{12}^{-1}r_{13}^{-1}$ and the $3^{rd}$ is $\binom{N}{2}\binom{N-2}{2} <y_0|r_{12}^{-1}r_{34}^{-1}|y_0>\approx \binom{N}{2}\binom{N-2}{2}N^{-4} \int_4(\Pi_{i=1}^4\rho_0(\mathbf{r}_i,a))r_{12}^{-1}r_{34}^{-1}= ((N-1)(N-2)(N-3)/(4N^3))\int_4(\Pi_{i=1}^4\rho_0(\mathbf{r}_i,a))r_{12}^{-1}r_{34}^{-1} = ((N-1)(N-2)(N-3)/(4N^3)) [\int_2\rho_0(\mathbf{r}_1,a)\rho_0(\mathbf{r}_2,a)r_{12}^{-1}]^2 \approx (¼) [\int_2\rho_0(\mathbf{r}_1,a)\rho_0(\mathbf{r}_2,a)r_{12}^{-1}]^2$, finally,

$$<y_0|H_{ee}^2|y_0>\approx (½)\int_2(\Pi_{i=1}^2\rho_0(\mathbf{r}_i,a))r_{12}^{-2} + \int_3(\Pi_{i=1}^3\rho_0(\mathbf{r}_i,a))r_{12}^{-1}r_{13}^{-1} + (¼)[\int_2(\Pi_{i=1}^2\rho_0(\mathbf{r}_i,a))r_{12}^{-1}]^2. \quad (19)$$

The ½, 1 and ¼ approximate coefficients in Eq.19 are "elegant", but one should use their accurate algebraic function with N (the (N-1)/(2N), etc.) for improved accuracy, the DFT approx. (use of $\rho_0$ instead of $y_0$ starting from Eq.18) makes unavoidable inaccuracy, anyway. More, the accurate wave function form in Eq.17 and the approx. DFT form in Eq.19 need about the same programming and calculation time for single determinants $S_0$ (via a=1) and $Y_0$ (via a=0), so one should use Eq.17, the Eq.19 should be considered as an elegant DFT approx. only. The $3^{rd}$ term in Eq.19 can be back-substituted with wave function form in Eq.18 as

$$<y_0|H_{ee}^2|y_0>\approx (½)\int_2(\Pi_{i=1}^2\rho_0(\mathbf{r}_i,a))r_{12}^{-2} + \int_3(\Pi_{i=1}^3\rho_0(\mathbf{r}_i,a))r_{12}^{-1}r_{13}^{-1} + [<y_0|H_{ee}|y_0>]^2 \quad (20)$$

showing a certain power expansion, as well as the missing terms in relation to the weaker

$$\langle y_0|H_{ee}^2|y_0\rangle \approx [\langle y_0|H_{ee}|y_0\rangle]^2 \quad \text{along with} \quad \langle y_0|H_{ee}^2|y_0\rangle > [\langle y_0|H_{ee}|y_0\rangle]^2 > 0 . \quad (21)$$

All terms in Eqs.17-21 are positive and can be used in Eq.10. For example, Eq.20 vs. 21 is the source of error in moment approx. in Table 1 which uses the simple definitions through $w_2$ and $w_3$. Terms $0 < \langle y_0|(H_{ne}+H_{ee})^2|y_0\rangle = \langle y_0|H_{ne}^2|y_0\rangle + 2\langle y_0|H_{ne}H_{ee}|y_0\rangle + \langle y_0|H_{ee}^2|y_0\rangle$ coming up in Eq.11, i.e. $\langle y_0|H_{ne}^2|y_0\rangle > 0$ and $\langle y_0|H_{ne}H_{ee}|y_0\rangle < 0$, can be analyzed and evaluated in analogous way to Eqs.17-21. Eq.10 simplifies with Eqs.9 and 20 as $e_{electr,0}^2 + 2e_{electr,0}\langle Y_0|H_{ee}|Y_0\rangle + [Y_0|H_{ee}|Y_0\rangle]^2 + (\frac{1}{2})\int_2(\Pi_{i=1}^2\rho_0(\mathbf{r}_i,a=0))r_{12}^{-2} + \int_3(\Pi_{i=1}^3\rho_0(\mathbf{r}_i,a=0))r_{12}^{-1}r_{13}^{-1}$, that is

$$E_{electr,0}^2 \approx [E_{electr,0}(\text{TRNS with } H^2 \text{ trick})]^2 \approx [E_{electr,0}(\text{TRNS})]^2 + (\frac{1}{2})\int_2(\Pi_{i=1}^2\rho_0(\mathbf{r}_i,a=0))r_{12}^{-2} + \int_3(\Pi_{i=1}^3\rho_0(\mathbf{r}_i,a=0))r_{12}^{-1}r_{13}^{-1}. (22)$$

**TABLE 1**: Average deviation of $E_{electr,0}$(HF-SCF/6-31G*/a=1) and $E_{electr,0}$(HF-SCF/6-31G*/a=0 moment approx.) with 1st (same as Eq.9), 2nd (reduction from Eq.10) and 3rd level terms $c_3 z_0^{w3}$ from $E_{electr,0}$(CI by Davidson at al. [6]) for 185 atomic ions ($1 \leq N, Z_A \leq 18, M=1$, i.e. H, H⁻, He⁺, …, Ar⁺, Ar)

|  | $c_2$ | $c_3$ | $w_2$ | $w_3$ | Deviation from CI/ hartree |
|---|---|---|---|---|---|
| HF-SCF/6-31G*/a=1 | - | - | - | - | 0.3951 |
| 1st level moment | - | - | - | - | 1.7791 |
| 2nd level moment | 1.1556 | - | 2.0 | - | 0.5171 |
| 3rd level moment | 0.9668 | 0.162 | 1.999 | 0.838 | 0.6977 |

**TABLE 2**: Some configurations to set up single determinant basis set $\{S_k\}$ or $\{Y_k\}$ for (N=9, 2S+1=2 doublet, open shell) and (N=10, 2S+1=1 singlet, closed shell) using only LUMO ($i_{max}$=5) to generate excited determinants (k>0), e.g. for the considered Ar⁹⁺ and Ar⁸⁺, resp. The i=4, 5 is called HOMO, LUMO, resp. in both k=0 ground states. It is the generation of configurations for any molecule with this (N, 2S+1), for example, the singlet H₂O and CH₄ possessing N=10 electrons, etc. The energy calculation is simpler (see Eqs.5-6) for $\{Y_k\}$ than for $\{S_k\}$ and that is listed only. The dash (-), full and ↑ mean that the energy level is occupied by no, α and β opposite spins (↑↓) and α spin (↑), resp. Further configurations can be systematically generated by elementary combinatorics for both, N=9 and 10. The $\varepsilon_{i+1} > \varepsilon_i$ for all i=0…5 comes from Eq.2, where degenerate states ($\varepsilon_{i+1} = \varepsilon_i$ for some i) can come up. Important is that, in TRNS (a=0), the energy levels $\{\phi_i, \varepsilon_i\}$ via Eq.2 do not alter with configurations (Eqs.5-6) and N, (but it does if a≠0). Increasing $i_{max}$ can drastically increase the configurations [3], so choosing good quality, small number and (energetically) low lying determinants is vital for effective calculations for ground and a few lowest lying excited states. Notice that, "single excitation"= moving 1 electron, "singlet excitation"= in resulting determinant the 2S+1=1.

| Energy levels | Some configurations for N= 10 and 2S+1=2Σsᵢ+1= 1 (singlet excitation) | | | | | | Some configurations for N= 9 and 2S+1=2Σsᵢ+1= 2 (doublet excitation) | | | | |
|---|---|---|---|---|---|---|---|---|---|---|---|
| $\varepsilon_5, \phi_5$ LUMO | - | full | full | full | ↑ | ↓ | - | ↑ | ↑ | ↑ | ↑ |
| $\varepsilon_4, \phi_4$ HOMO | full | - | full | full | ↓ | ↑ | ↑ | - | full | full | full |
| $\varepsilon_3, \phi_3$ | full | full | - | full | full | full | full | full | - | full | full |
| $\varepsilon_2, \phi_2$ | full | full | full | - | full | full | full | full | full | - | full |
| $\varepsilon_1, \phi_1$ | full | full | full | full | full | full | full | full | full | full | - |
| $\varepsilon_0, \phi_0$ | full | full | full | full | full | full | full | full | full | full | full |
| $e_{electr,k}$ (where f≡ 2 $\sum_{i=0}^{5} \varepsilon_i$) | f-2$\varepsilon_5$ | f-2$\varepsilon_4$ | f-2$\varepsilon_3$ | f-2$\varepsilon_2$ | f-$\varepsilon_4$-$\varepsilon_5$ | f-$\varepsilon_4$-$\varepsilon_5$ | f-2$\varepsilon_5$-$\varepsilon_4$ | f-2$\varepsilon_4$-$\varepsilon_5$ | f-2$\varepsilon_3$-$\varepsilon_5$ | f-2$\varepsilon_2$-$\varepsilon_5$ | f-2$\varepsilon_1$-$\varepsilon_5$ |
| excitation | no | dbl | dbl | dbl | single | single | no | single | dbl | dbl | dbl |
| k in $S_k$ or $Y_k$ | 0 | 1 | 2 | 3 | 4a | 4b | 0 | 1 | 2 | 3 | 4 |

## ACKNOWLEDGMENTS

Financial and emotional support for this research from OTKA-K 2015-115733 and 2016-119358 are kindly acknowledged. Special thanks to Hermin Szeger for her help in typing the manuscript. The subject has been presented in ICNAAM_2019, Greece, Rhodes and published in AIP 2020.